\documentstyle[12pt,aaspp4]{article}
\begin{document}

\newcommand{\up}[1]{\ifmmode^{\rm #1}\else$^{\rm #1}$\fi}
\newcommand{\zdot}{\makebox[0pt][l]{.}}
\newcommand{\upd}{\up{d}}
\newcommand{\uph}{\up{h}}
\newcommand{\upm}{\up{m}}
\newcommand{\ups}{\up{s}}
\newcommand{\arcd}{\ifmmode^{\circ}\else$^{\circ}$\fi}
\newcommand{\arcm}{\ifmmode{'}\else$'$\fi}
\newcommand{\arcs}{\ifmmode{''}\else$''$\fi}

\title{Photometric standard stars in the BVI system in a wide
field centered on the spiral galaxy NGC 300.\footnote{Based on  
observations obtained with the 1.3-m
Warsaw telescope at the Las Campanas  Observatory of the Carnegie
Institution of Washington and with the 2.2-m ESO/MPI telescope at the 
European Southern Observatory}}

\author{G. Pietrzy{\'n}ski}
\affil{Universidad de Concepci{\'o}n, Departamento de Fisica,Casilla 160--C, 
Concepci{\'o}n, Chile}
\affil{Warsaw University Observatory, Al. Ujazdowskie 4,00-478, Warsaw}
\authoremail{pietrzyn@hubble.cfm.udec.cl}
\author{W. Gieren}
\affil{Universidad de Concepci{\'o}n, Departamento de Fisica,Casilla 160--C, 
Concepci{\'o}n, Chile}
\authoremail{wgieren@coma.cfm.udec.cl}
\author{A. Udalski}
\affil{Warsaw University Observatory, Al. Ujazdowskie 4,00-478, Warsaw}
\authoremail{udalski@astrouw.edu.pl}

\begin{abstract}
Based on 13 nights of observations of four fields in NGC 300, 
we have set up an extensive sequence of stars with accurate BVI photometry
covering a relatively large (25 x 25 arcmin) region centered on this galaxy. 
This
sequence of standard stars is very  useful for calibrating the photometry of 
variable stars and other objects in NGC 300 and other galaxies obtained
from wide field mosaic images.
Our standard star list contains B, V and I measurements  for 390 stars. 
The accuracy of the zero points in the V filter and B-V color is better than
0.02 mag, and about 0.03 mag for the V-I color. We found very good 
agreement between our measurements and those
previously obtained by Walker for 26 stars near NGC 300.  

\end{abstract}


\section{Introduction}

We are currently engaged in a detailed study of the stellar populations in the 
southern spiral galaxy NGC 300. This object belongs to the Sculptor Group and 
is, at a distance of about 2 Mpc, one of the closest spirals, close enough 
to be resolved into stars even near the nucleus on high-resolution
ground-based images.  As a first step, we recently investigated the OB 
associations content of NGC 300 (Pietrzy{\'n}ski et al. 2001) from wide field 
images obtained at the ESO 2.2-m telescope, which cover a 34 x 34 arcmin field 
centered on the galaxy. Based on the same data 117 Cepheids and
12 Cepheid candidates were found (Pietrzy{\'n}ski et al. 2002),
 extending the previous work of Graham (1984) who discovered the first 
Cepheids in NGC 300 from photographic plates, and Freedman et al. (1992) who 
derived the first CCD light curves of 12 of these Cepheids, which were used to 
derive a preliminary distance to the galaxy. 
The newly discovered Cepheids will be very useful to obtain a truly accurate 
distance to this 
galaxy, and to calibrate the effect of metallicity on the Cepheid 
period-luminosity relation in optical (BVRI) bandpasses.
In a parallel project, we analyzed  photometric and spectroscopic data for a 
sample of 
blue supergiants in NGC 300 with the aim to obtain metal abundances and the
metallicity gradient in the disk of NGC 300, and to derive an independent 
distance
estimate from the wind momentum-luminosity relation valid for these stars
(Kudritzki et al. 1999). First results of this study have been recently 
presented by Bresolin et al. (2002). 

In order to accurately calibrate the photometric magnitudes of the Cepheids, 
blue supergiants,
and of thousands of other  stars in our wide field images, we need a sufficient 
number of 
high-quality photometric standard stars which are distributed over the whole
 area we are observing, span a broad range in magnitudes and colors, and 
extend down to the faint limit of our 1.3 m images. A previous photometric sequence 
close to NGC 300 has been set up by Graham (1981; photoelectric sequence 
covering the magnitude range V=9-20), which was later improved by Walker 
(1988, 1995) from CCD photometry. However, this sequence is covering only a
 relatively small field (about 6 x 6 arcmin) and contains 
only a small number of stars with 
BVI magnitudes (26). This is insufficient for our purpose to provide highly 
accurate (to 0.02-0.03 mag) BVI magnitudes for thousands of stars in the 
wide 34 x 34 arcmin field
around NGC 300. We therefore decided to set up a new and extensive sequence 
of accurate standard star magnitudes in this field, well suited for the
transformation of data from wide field, mosaic detectors. 

\section{Observations}

We have collected  images of NGC 300 through standard BVI filters using the 
very stable photometric telescope/detector system of the Warsaw 1.3 m
 telescope at Las Campanas Observatory, which has been used to conduct the 
OGLE II photometric microlensing survey (Udalski, Kubiak, and Szyma{\'n}ski 
1997).
The telescope was equipped with a 2048 $\times$ 2048 CCD detector. The pixel
 size was 24 $\mu$m, which corresponds to a scale of 0.417 arcsec/pixel. The
 observations were performed in the "medium" reading mode of the CCD detector.
 The gain and readout noise were 7.1 electrons/ADU and 6.3 electrons, 
respectively. More details about the instrumentation setup can be found in 
Udalski,
 Kubiak and Szyma{\'n}ski (1997). Color equations, and the transformations to 
the BVI (Johnson-Cousins) standard photometric system are very  well 
established for this system, making it an ideal choice to carry out our 
programme of 
setting up a new standard star sequence in NGC 300. In order to cover  
most of the  34 x 34 arcmin field covered by our ESO wide field images, we 
observed the 
four slightly overlapping 15 x 15 arcmin fields shown in Fig. 1. Their 
coordinates are given in Table 1. 
During 13  nights in August, September and October 2000 
we were able to secure 48 images of these fields (16 in each filter). 
 The journal of these observations is given in Table 2. Exposure times 
were 900, 750, and 600  s in B, V, and I bands, respectively.

\begin{deluxetable}{c c c c}
\tablecaption{Observed fields in NGC 300}
\tablehead{
\colhead{Field} & \colhead{RA (J2000)} & \colhead{DEC (J2000)} & 
\colhead{$N_{nights}$}
}
\startdata
F1   &0\uph54\upm21\zdot\ups2 &-37\arcd34\arcm40\arcs&5\\
F2  &0\uph54\upm21\zdot\ups2  &-37\arcd48\arcm04\arcs&4\\
F3 &0\uph55\upm24\zdot\ups6   &-37\arcd48\arcm57\arcs&4\\
F4  &0\uph55\upm24\zdot\ups6  &-37\arcd34\arcm34\arcs&3\\

\enddata

\end{deluxetable}

\begin{deluxetable}{c c c c}
\tablecaption{Journal of observations}
\tablewidth{0pt}
\tablehead{
\colhead{Night (2000)} & \colhead{Field(s)} & \colhead{Filters} & \colhead{seeing 
(arcsec)} 
}
\startdata
Aug02 & FI & BVI & 1.0  \\
Aug05 & FII & BVI & 1.1  \\
Aug06 & FIII & BVI & 0.9  \\
Aug07 & FIV & BVI & 0.8  \\
Aug30 & FI,FII& BVI & 1.2  \\
Sep15 & FI& BVI & 1.1   \\
Sep22 & FIII,FIV & BVI & 0.7  \\
Sep25 & FI & BVI  &1.0  \\
Sep26 & FII & BVI & 1.2  \\
Sep27 & FIII& BVI & 1.3  \\
Sep28 &FI,FIV & BVI  & 1.3  \\
Sep30 &FII & BVI  & 0.9  \\
Oct03 & FIII & BVI &1.2  \\
\enddata
\end{deluxetable}

\section{Reductions}
The raw frames have been de-biased and flatfielded in the standard way using the
IRAF\footnote{IRAF is distributed by the
National Optical Astronomy Observatories, which are operated by the   
Association of Universities for Research in Astronomy, Inc., under cooperative
agreement with the NSF.} package. Because of the variations of the Point Spread 
Function (PSF) across the image we divided our frames into four overlapping 
subframes, on which the PSF can be reasonably assumed as being constant. Then 
profile photometry with the DAOPHOT and ALLSTAR programs  was carried out on each of 
the subframes. The  PSF model was derived iteratively. First 15 candidates
 from relatively bright and isolated stars were selected and the first 
approximation of the PSF was derived using DAOPHOT. In the next step 
we subtracted all neighbouring stars with the ALLSTAR program and derived a new 
PSF.  After three such  loops no improvement to the PSF  was noted anymore, and we 
obtained in this way the PSF model which we finally adopted.

In order to convert our profile photometry to the aperture system,
aperture corrections were derived for each of the four subframes of any
given frame. For this purpose a special procedure was prepared based on DAOPHOT.
 First the bright, well separated stars were
selected (usually 10 per subframe). Then iteratively all neighbouring
stars, which could contaminate our measurements  were removed
and aperture photometry was obtained. Finally, the median of the aperture 
corrections obtained for these stars was adopted as the
aperture correction of the given subframe. With the aperture corrections
for all our subframes we transformed the profile photometry to the
aperture system and merged them into one file. Relatively large   
overlapping fields (about 400 pixels) allowed us to check in detail the           
consistency of our procedure of determining the aperture corrections.
Usually more than 20 stars with good photometry (e.g. well suited
for comparison purposes) were found to be common to a given two subframes.
The mean difference between their aperture photometric magnitudes did not
exceed 0.005 mag and the rms was smaller than 0.007 mag. No dependence 
on brightness, position or color was found.

\section{Transformations}

\subsection{Transformation coefficients}
During 10 nights between 3 and 7 Landolt (1992) standard star fields were observed in 
order to 
transform our instrumental magnitudes to the standard $BVI$  system. We adopted 
the following transformations: 

$$B=b-0.041\times (B-V) + {\rm const}_B$$
$$V=v-0.002\times (V-I) + {\rm const}_V$$
$$I=i+0.029\times (V-I) + {\rm const}_I \eqno{(1)}$$
$$B-V= 0.959\times (b-v) + {\rm const}_{B-V}$$
$$V-I= 0.969\times (v-i) + {\rm const}_{V-I}$$

\noindent where the lower case letters {\it b,v,i} denote the aperture 
instrumental 
magnitudes
normalized to 1 sec exposure time. The color coefficients we used were 
established 
during previous extensive observations of a
large number of standard stars over the entire season by the OGLE team
(e.g. Udalski et al. 1998, 2000). The extinction coefficients were derived 
for 7 nights. For the other nights the mean extinction coefficients 
were used.
The zero points were determined for all of these 10 nights.
The residuals  did not exceed 0.03 mag and the zero point was very 
stable during the whole run (see Fig. 2). The residuals did not show any 
significant dependence on color, brightness or  position on the sky.
For remaining 3 nights we assumed  mean zero points derived during 
our run.

In order to test the mean color coefficients provided by the OGLE team we 
observed,
during three nights,  a large number of standard stars covering a large 
range in air mass and color and derived the full set of transformation 
coefficients from these observations. Then we transformed  the instrumental 
magnitudes using 
these coefficients and mean ones, and compared the results. The comparison 
showed that the magnitude difference was smaller than 0.008 mag in each band, 
demonstrating the high accuracy and stability 
of the adopted transformation equations.

\subsection{Astrometric transformation} 
To convert pixel positions of stars to the equatorial coordinate system we used 
the 
algorithm developed and used in the OGLE project. In brief, using the 
Digital Sky Survey (DSS) images, fits files slightly larger than our
observed fields were extracted. All stars having 200 counts above 
sky level were detected on these frames and their centroids were 
calculated. Then, 
(x,y) pixel coordinates of each star from the DSS images were converted 
to equatorial coordinates. The resulting  (RA,DEC) of our stars were then 
transformed 
to (x',y') pixel coordinates on the plane tangent to the celestial 
sphere at the center of a given field. Finally, the two sets of pixel 
coordinates 
(x,y) and (x',y') were tied together using simple third order polynomials.
To check the consistency of our transformation we examined the derived 
coordinates 
for common stars from overlapping regions. The mean difference of the 
coordinates 
derived from the different transformations was found to be about 0.3 arcsec.  

\section{Results}
\subsection{BVI photometry}
In order to derive final magnitudes and colors, the instrumental magnitudes were 
 normalized to 1 s exposure time
and tied to the aperture system (see Section 3). Then they were transformed 
to the standard BVI system using the coefficients derived as
described in Section 4.1.
For stars from  each of the four fields the mean, 
and the standard deviations were calculated from all observations 
(usually 3-5 measurements per star). Fig. 3 presents the standard deviations 
in B, V and I bands, as a function of magnitude. It is appreciated that 
there are  many stars with relatively small scatter (i.e. good
candidates for being local standard stars).
 To check on the consistency of our photometric data 
from the 4 different fields we used stars from the overlapping regions. 
More details of the comparison of the magnitudes and colors of common stars
in the four overlapping regions can be found in Table 3.
It is seen that the mean differences were usually smaller than 0.02
mag, with a comparable scatter, which indicates that there is no systematic
difference between the photometry for our four fields. One should note that the
majority of the observations of the four regions were performed during
different nights.

\begin{deluxetable}{c c c c c c c c c c c c}
\tablecaption{Results of a comparison of 
common stars in four overlapping regions}
\tablewidth{0pt}
\tablehead{
\colhead{Region}& N & \colhead{ $<\delta_{\rm B}>$ } & \colhead{ 
$\sigma_{\delta_{\rm B}}$} &
\colhead{ $<\delta_{\rm V}>$ } & \colhead{ $\sigma_{\delta_{\rm V}}$ }  &
\colhead{ $<\delta_{\rm I}>$ } & \colhead{ $\sigma_{\delta_{\rm I}}$} & 
\colhead{$<\delta_{\rm BV}>$} & \colhead{$\sigma_{\delta_{\rm BV}}$} &
\colhead{$<\delta_{\rm VI}>$} & \colhead{$\sigma_{\delta_{\rm VI}}$}
}

\startdata
FI-FII & 26 & -0.011  & 0.019 & 0.000  & 0.035 & -0.027 & 0.018  & 0.002 & 0.014 & 
0.004 &0.022 \\
FI-FIV &21 & -0.010  & 0.026 & 0.003  & 0.016 & -0.016 & 0.017  & -0.010 &0.024 & 
0.012 &0.021 \\
FII-FIII &19 & 0.013  & 0.035 & 0.006  & 0.030 & -0.001 & 0.026  & 0.019 &0.018 & 
-0.015 &0.028\\
FIII-FIV &25& -0.003  & 0.022 & 0.019  & 0.020 & 0.028 & 0.023  & -0.023 & 0.019 & 
-0.007 &0.024\\
\enddata

\end{deluxetable}

\subsection{Comparison with previous results}
In order to compare our photometry with that obtained by Walker (1995)
for 26 stars in the vicinity of NGC 300, we identified 17 of these stars 
in our database. Remaining stars from Walker's list were too bright or situated 
outside our observed field.

 The mean difference between Walker's photometry and
ours is ${\Delta V=0.002\pm0.025}$~mag, ${\Delta(B-V)=-0.008\pm0.025}$~mag
and ${\Delta(V-I)=-0.019\pm0.027}$~mag. Evidently the results from both studies 
are
in excellent agreement.

Freedman et al. (1992) have obtained CCD BVRI observations for 16 Cepheids in 
NGC 300.
 The comparison of our new data on these Cepheids to the 
Freedman et al. data is presented in 
Pietrzy{\'n}ski et al. (2002), and the reader is referred to that paper for more 
details. Here we only note that very good agreement 
was found. The intensity mean magnitudes of the common Cepheids usually  did not 
differ by more than 0.05 mag. 
Differences between individual observations at similar phases from the 
two discussed data sets 
were typically $\pm$ 0.03-0.05 mag (see Fig. 6 in Pietrzy{\'n}ski et al. 2002). 

All these comparisons make us conclude that our photometry is very  
well tied to the standard system. Especially the very small systematic  difference 
between our photometry and the very well calibrated  sequence of standard 
stars of Walker is gratifying. Unfortunately all stars observed by Walker (1995) were located 
in only one (F II) of our four fields. Taking into account that typically the differences 
in photometry between  our fields are about 0.01 mag (see Table 3) we estimate that 
our new standard star magnitudes are accurate to better than 0.02 mag in V and B-V, 
and about 0.03 mag in the case of V-I.   

\subsection{BVI photometric sequence}
The main goal of this paper is the presentation of an extensive list of
secondary standard stars covering a large area around NGC 300, 
well suited for the absolute calibration of data obtained with 
wide field, mosaic detectors. Having a large number of stars with accurate 
magnitudes and colors over the whole area of a given mosaic detector (usually 
the field of view is about 0.5 x 0.5 square degrees) one can easily 
transform the observations from all the chips  into the standard system, 
avoiding problems with possible differences in the color coefficients, zero 
points 
etc.  between the various CCDs of the mosaic camera. 

To filter out appropriate standard stars, we need to exclude spurious 
stars detected in the wings of saturated objects, stars with their photometry 
affected by saturation, stars too close to the edges of the frames,  
stars being seriously blended, non-stellar objects, and possible variable 
stars.
In order to weed out any such potential problem stars from our sample, we 
applied
the following selection criteria:

\noindent 1) A star must have at least 2  measurements \\
2) Standard deviation in V, I and V-I  $<$0.05 mag, in B and B-V $<$
0.07 mag \\ 
3) Distance from the edge of a chip must be at least 10 arcsec. \\
4) Counts must be less than  90 \% of the detector range. \\

As a next step, we visually examined the light curves for all the  stars passing 
the selection criteria
using a database constructed from 29 nights of observations of 
NGC 300 with the ESO WFI imager. All stars showing evidence for 
possible variability were rejected.

Alltogether 390 stars passed our selection process  and entered into 
the final list of standards which is presented in Table 4.
 These stars span the following range in brightness 
and colors: 14.2 $<$ V $<$ 22.0, -0.35 $<$ B-V $<$ 1.94, -0.35 $<$ V-I $<$ 3.26.
Information about typical standard deviations for our standard stars in different 
bands as a function of V magnitude is given in Table 5.

\setcounter{table}{4}

\begin{deluxetable}{c c c c c}
\tablecaption{Mean standard deviations of standard stars 
as a function of their V magnitude.}
\tablewidth{0pt}
\tablehead{
\colhead{Magnitude range in V} & \colhead{$<\sigma_{\rm B}>$} & \colhead{$<\sigma_{\rm V}>$} &
\colhead{$<\sigma_{\rm I}>$} & \colhead{N stars} \\
}
\startdata
14-15 &   0.014 &   0.012 &   0.013 & 10 \\ 
15-16 &   0.016 &   0.014 &   0.014 & 14 \\ 
16-17 &   0.015 &   0.016 &   0.015 & 21 \\ 
17-18 &   0.018 &   0.014 &   0.013 & 33 \\ 
18-19 &   0.018 &   0.016 &   0.015 & 49 \\ 
19-20 &   0.020 &   0.018 &   0.018 & 100 \\ 
20-21 &   0.022 &   0.020 &   0.021 & 118 \\ 
21-22 &   0.029 &   0.027 &   0.023 & 45 \\ 
\enddata
\end{deluxetable}

\section{Summary}

We have obtained accurate BVI photometry for a large number of stars located in 
a relatively large 
(25 x 25 arcmin) field centered on NGC 300. The photometric zero 
points are accurate to 0.02 mag in V and B-V and to 0.03 mag in V-I. 
From our database, we have selected the
stars with the most reliable and accurate photometry which form our 
final catalog of 390  secondary standard stars in NGC 300. 
These stars cover a very large range in brightness and color. We have already 
used these standard stars
 for our recent photometric work on the stellar populations in NGC 300, 
and we believe that the new catalog will be very useful for researchers 
observing
this particular galaxy or any other objects in this part of the sky with modern 
wide 
field CCD mosaic detectors.

We intend to continue our observations of NGC 300 in order to improve
and extend our list of secondary standard stars in this field.

\acknowledgments
We would like to thank the OGLE team, especially Drs: 
Michal Szyma{\'n}ski and Marcin Kubiak for their kind help with the observations
and data reduction. 
We acknowledge the use of The Digitized Sky Survey which was produced at the 
Space Telescope Science Institute based on photographic data obtained with 
The UK Schmidt Telescope, operated by the Royal Observatory Edinburgh. WG 
gratefully acknowledges financial support for this work received from Fondecyt 
Lineas Complementarias grant 8000002. The paper was partly supported
from the KBN BST grant for Warsaw University Observatory. We also would 
like to thank the anonymous referee for interesting comments and suggestions.

\newpage

\begin{figure}[htb]
\caption{Observed fields in NGC 300. Displayed region corresponds to 
about 34 x 34 arcmin on the sky. North is up and east to the left.}
\end{figure}  

\begin{figure}[htb]
\vspace*{12 cm}
\includegraphics{pietrzynski.fig2.ps}
\caption{The variation of the photometric zero
points in the B,V and I filters over our observing nights}
\end{figure}

\begin{figure}[htb]
\vspace*{10 cm}
\includegraphics{pietrzynski.fig3.ps}
\caption{Standard deviation of brightness as a
function of magnitude, in different bands, for all observed stars in NGC 300}
\end{figure}

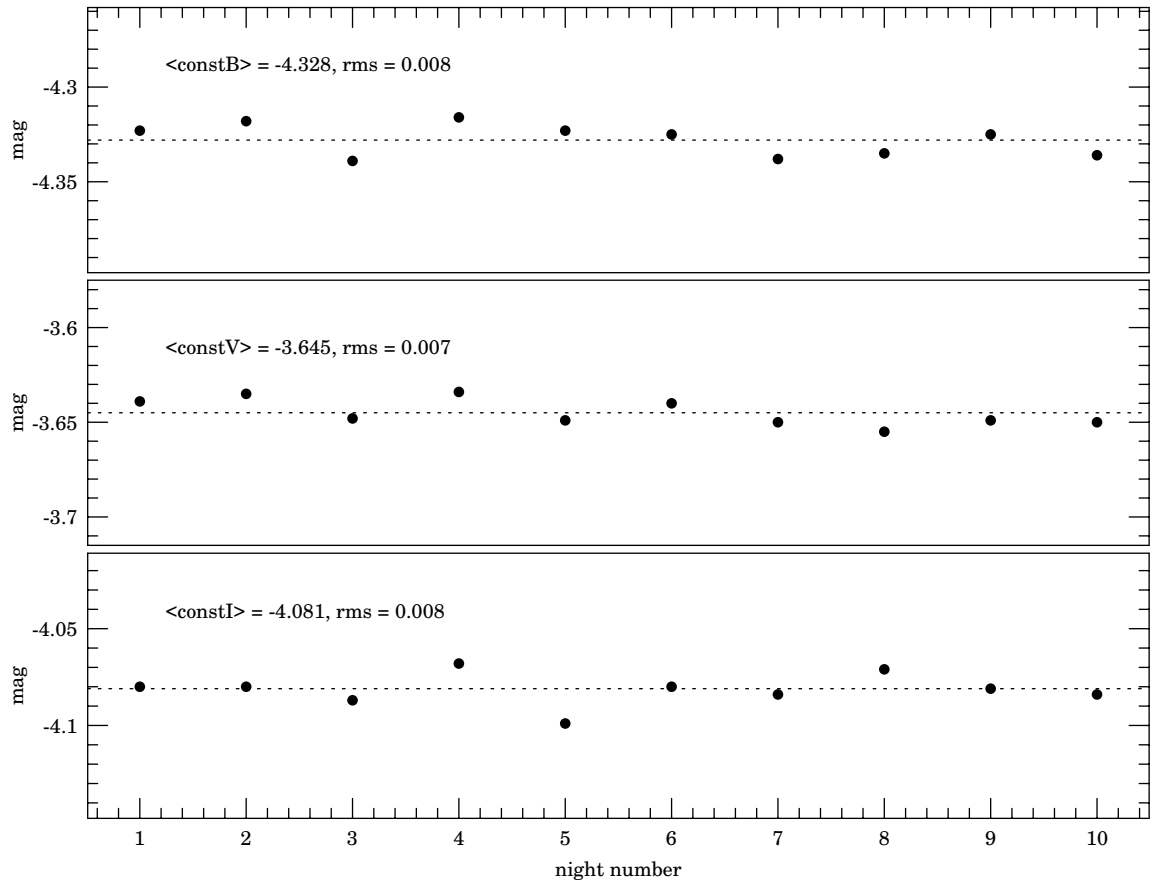
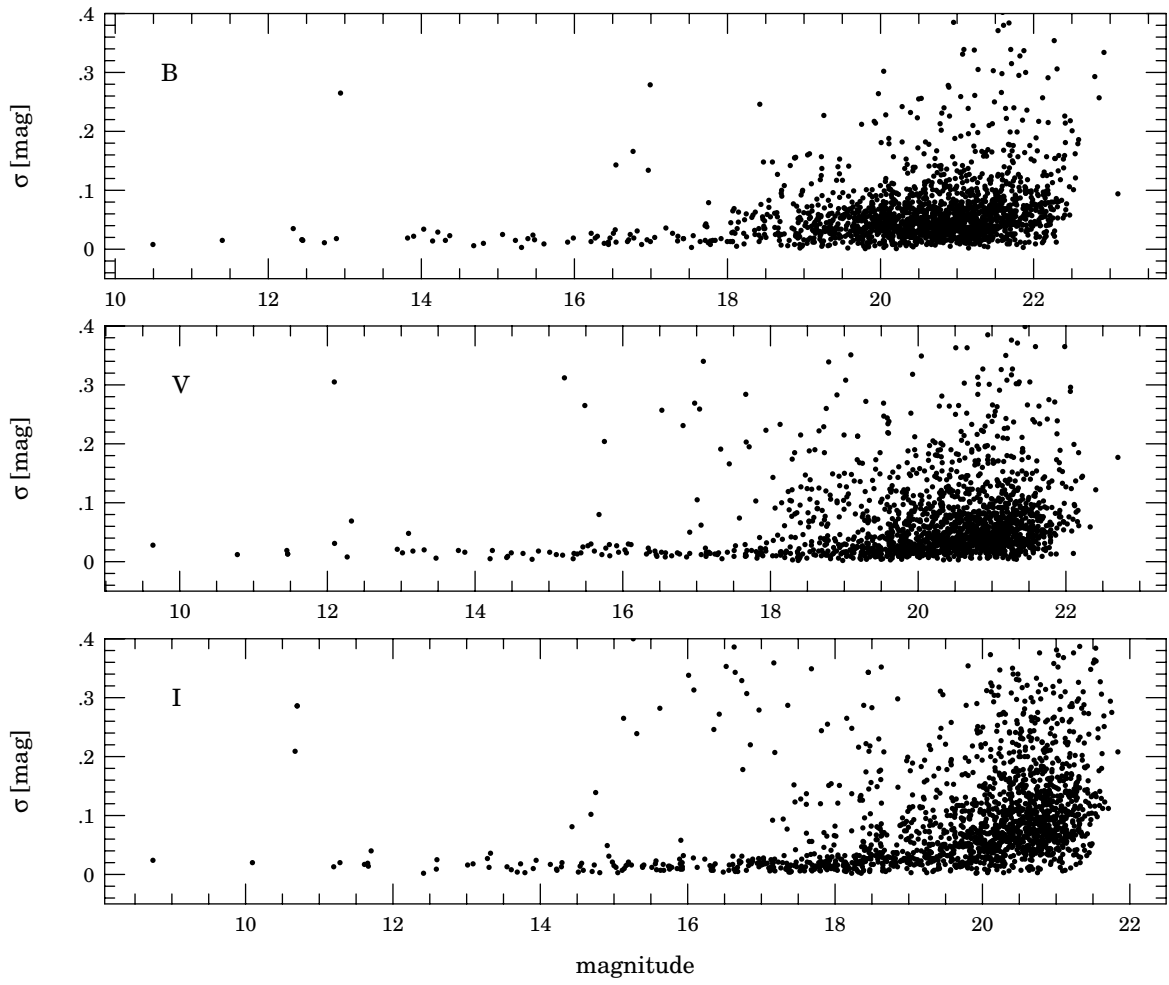
\end{document}